\documentstyle[12pt,a4]{article}
\title{The Mathai-Quillen Formalism and Topological Field Theory%
       \thanks{Notes of lectures given at the Karpacz Winter
               School on `Infinite Dimensional Geometry in Physics'
               (17 - 27 February 1992).}}

\author{Matthias Blau%
        \thanks{e-mail: t75@nikhefh.nikhef.nl, 22747::t75} \\
        NIKHEF-H                                           \\
        P.O. Box 41882, 1009 DB Amsterdam                   \\
        The Netherlands}

\date{March 9, 1992}

\begin{document}
\newcommand{\nc}{\newcommand}
\newcommand{\rnc}{\renewcommand}
\catcode`\@=11                                         
\@addtoreset{equation}{section}                        
\rnc{\theequation}{\arabic{section}.\arabic{equation}} 
\nc{\be}{\begin{equation}}
\nc{\ee}{\end{equation}}
\nc{\bea}{\begin{eqnarray}}
\nc{\eea}{\end{eqnarray}}
\rnc{\a}{\alpha}
\nc{\g}{\gamma}
\rnc{\d}{\delta}
\nc{\e}{\eta}
\nc{\eb}{\bar{\eta}}
\nc{\f}{\phi}
\nc{\fb}{\bar{\phi}}
\nc{\vf}{\varphi}
\nc{\p}{\psi}
\rnc{\pb}{\bar{\psi}}
\rnc{\c}{\chi}
\nc{\cb}{\bar{\c}}
\nc{\m}{\mu}
\nc{\n}{\nu}
\rnc{\o}{\omega}
\rnc{\t}{\theta}
\nc{\tb}{\bar{\theta}}
\nc{\M}{{\cal M}}                           
\nc{\C}{{\cal A}/{\cal G}}                  
\nc{\A}[1]{{\cal A}^{#1}/{\cal G}^{#1}}     
\nc{\RC}{{\cal R}_{\C}}                     
\nc{\RM}{{\cal R}_{\M}}                     
\nc{\RX}{{\cal R}_{X}}
\nc{\RY}{{\cal R}_{Y}}
\nc{\ra}{\rightarrow}
\nc{\ot}{\otimes}
\rnc{\ss}{\subset}
\rnc{\lg}{{\bf g}}                          
\nc{\cs}{\c_{s}}                            
\nc{\del}{\partial}
\nc{\dx}{\dot{x}}
\nc{\F}{\Phi}
\nc{\Fn}{\Phi_{\nabla}}
\nc{\On}{\Omega_{\nabla}}
\nc{\tint}[1]{\int_{0}^{#1}\!dt\;}
\nc{\nt}{\nabla_{t}}
\def\s1{S^{1}}
\nc\en{e_{\nabla}}
\nc\ens{e_{s,\nabla}}
\nc\env{e_{V,\nabla}}
\rnc{\O}[2]{\Omega^{#1}({#2},\lg)}          
\def\tft{top\-o\-log\-i\-cal field the\-o\-ry}
\def\tgt{top\-o\-log\-i\-cal gauge the\-o\-ry}
\def\sqm{su\-per\-sym\-met\-ric quan\-tum mech\-an\-ics}
\def\mq{Mathai-Quillen}
\maketitle

\begin{abstract}
These lecture notes give an introductory account of an approach to
cohomological field theory due to Atiyah and Jeffrey which is based
on the construction of Gaussian shaped Thom forms by Mathai and Quillen.
Topics covered are: an explanation of the Mathai-Quillen formalism
for finite dimensional vector bundles; the definition of regularized
Euler numbers of infinite dimensional vector bundles; interpretation
of supersymmetric quantum mechanics as the regularized Euler number of
loop space; the Atiyah-Jeffrey interpretation of Donaldson theory;
the construction of topological gauge theories from infinite dimensional
vector bundles over spaces of connections.
\end{abstract}
\vfill
\begin{flushright}
NIKHEF-H/92-07
\end{flushright}
\newpage

\tableofcontents

\section{Introduction}

Topological field theory has been a lively area for research ever since
the appearance of the seminal work by Witten \cite{ewdon,ewsig,ewcs}
a few years ago. Activity in the field increased when the observation
was made \cite{ewtg,distler} that topological gravity in two dimensions
is closely related to two-dimensional quantum gravity and its description
in terms of random matrix models. Several reviews of the subject are now
available\footnote{See \cite{dvv,dijk,ewint} for an account of the
relation among topological gravity, matrix models, intersection theory
on moduli space, and integrable models, and \cite{pr} for a general
review of \tft.}.

I will try to complement these existing reviews by focussing
on an approach to \tft\ based on the construction
by Mathai and Quillen \cite{mq} of Gaussian shaped Thom forms
for finite dimensional vector bundles. This very elegant approach is due
to Atiyah and Jeffrey \cite{aj} who realized that \tft\ could be
regarded as an infinite dimensional generalization of this construction.
There are several advantages of adopting this point of view. First
of all, it provides an {\em a priori} explanation of the fact that
finite dimensional topological invariants can be represented by
functional integrals, the hallmark of \tft. Moreover,
it has the charming property of giving a
unified description of all kinds of (cohomological) topological field
theories and \sqm. This has the added bonus of making this approach quite
elementary as it allows one to develop the main ideas in a quantum mechanical
setting and to then transfer them almost verbatim to field theory. Lastly,
it also provides some insight into the mechanism of the localization of
path integrals in \sqm\ and \tft.

To those already familiar with the subject, these lectures will hopefully
provide a new and perhaps enlightning  perspective on \tft. At the
same time they should, ideally, constitute an elementary introduction to the
subject requiring no prior knowledge of the field and little more than some
basic differential geometry and the ability to perform Gaussian integrals.

The recurrent theme in these notes will be the Euler
number of a vector bundle. In order to understand the basic idea of
the Atiyah-Jeffrey approach, let us therefore recall that classically
there exist two quite different prescriptions for calculating the
Euler number $\c(X)\equiv\c(TX)$ of (the tangent bundle of) a manifold $X$.
The first is topological in
nature and instructs one to choose a vector field $V$ on $X$ with isolated
zeros and to count these zeros with signs (this is the Hopf theorem).
The second is differential
geometric and represents $\c(X)$ as the integral over $X$ of a density
(top form) $\en$ constructed from the curvature of some connection
$\nabla$ on $X$ (the Gauss-Bonnet theorem). Likewise, the Euler number
$\c(E)$ of some other vector bundle
$E$ over $X$ can be determined in terms of either a section $s$ of $E$
or a connection $\nabla$ on $E$.

A more general formula, obtained by Mathai and Quillen \cite{mq},
interpolates between these two classical prescriptions. It relies on
the construction of a form $\ens(E)$ which depends on both a section $s$
and a connection $\nabla$. This form has the property that
\[ \c(E)=\int_{X}\ens(E)\]
for all $s$ and $\nabla$. Moreover, this equation reduces to the Hopf or
Gauss-Bonnet theorem for appropriate choice of $s$ (for isolated zeros
to the former and to the latter for $s=0$).

What Atiyah and Jeffrey \cite{aj} pointed out
was that, although $\en$ and $\int_{X}\en$ do not make sense for
infinite dimensional $E$ and $X$, the \mq\ form $\ens$ can be
used to formally define {\em regularized Euler numbers} $\cs(E)$
of such bundles by
\[ \cs(E):=\int_{X}\ens(E) \]
for certain choices of $s$. Although not independent of $s$, these
numbers $\cs(E)$ are naturally associated with $E$ for natural
choices of $s$ and are therefore likely to be of topological interest.

It is precisely such a representation of topological invariants (in
a non-technical sense\footnote{What is meant by
`topological' in this context is the invariance of numbers like
$\cs(E)$ under deformations of certain of the data entering into
its calculation. It is in this sense that the Donaldson invariants
of four-manifolds \cite{don}, which arise as correlation functions
of the field theory considered in \cite{ewdon}, are topological as
they are independent of the metric which enters into the definition
of the instanton moduli space. They are, however, not topological
invariants in the mathematical sense as they have the remarkable
property of depending on the differentiable structure of the four-manifold.})
by functional integrals which is the characteristic
property of topological field theories, and which could also be taken
as their definition. This suggests, that certain topological
field theories can be interpreted or obtained in this way.
It will be the main aim of
these notes to explain that this is indeed the case for the cohomological
theories (i.e.~not Chern-Simons theory and its siblings). The models
we consider explicitly are, in addition to \sqm, Donaldson theory \cite{ewdon}
and various theories of flat connections discussed e.g.~in
\cite{ewtop,bbt,ms} and \cite{gtlec,btmq,btqm}.
This framework is, however, broad enough to include topological sigma models,
twisted minimal models, and their coupling to topological gravity as well
(see \cite{ewwzw,aspin}).

The following notes consist of three sections, dealing with the
Mathai-Quillen formalism, \sqm, and \tgt\ respectively. Each section
begins with a brief review of the required mathematical background.
Thus section 2.1 recalls the classical expressions for the Euler class
and Euler number of a finite dimensional vector bundle. For our
present purposes the Euler number of a vector bundle is best
understood in terms of its Thom class and section 2.2 exlains this
concept. It also contains the construction of the Gaussian shaped
Thom form of Mathai and Quillen and its descendants $\ens$.
Section 2.3 deals with the
application of the Mathai-Quillen formalism to infinite dimensional
vector bundles and their regularized Euler number and introduces
the examples to be discussed in more detail in the subsequent sections.

Section 3.1 contains the bare essentials of the geometry of the loop space
$LM$ of a manifold $M$ necessary to apply the \mq\ formalism to its
tangent bundle. Section 3.2 exlains how \sqm\ can be interpreted as
defining or arising as a path integral representation of the
regularized Euler number of $LM$. Some related results like the
path integral proofs of the Gauss-Bonnet and Poincar\'e-Hopf theorems
are reviewed in the light of this derivation. In section 3.3 it is
shown that the finite dimensional \mq\ form can, in turn, be derived
from \sqm.

Section 4.1 deals with the geometry of gauge theories. We derive
an expression for the curvature form of the principal fibration
${\cal A}\ra\C$ and give a
formula for the Riemann curvature tensor of moduli subspaces $\M\ss\C$.
We also introduce those infinite dimensional bundles which will
enter into the subsequent discussion of \tgt.
In section 4.2 it is shown that the partition function of Donaldson
theory can be interpreted as the regularized Euler number of a
bundle of self-dual two-forms over $\C$. It also contains a
brief discussion of some properties of topological field theories
in general, as well as some remarks on the interpretation of
observables in the present setting.
Topological gauge theories of flat connections in two and three
dimensions are the subject of section 4.3.
In particular, in $3d$ we sketch the construction of a \tgt\ representing
the Euler characteristic of the moduli space of flat connections;
once directly from the tangent bundle of $\C$
and once from \sqm\ on $\C$. We also construct a two-dimensional
analogue of Donaldson theory representing intersection theory on
moduli spaces of flat connections.

The basic references for section 2.1 and 2.2 are Bott and Tu \cite{botu}
and Mathai and Quillen \cite{mq}. For section 2.3 see \cite{aj} and
\cite{btmq}. The main result of section 4.2 is due to Atiyah and Jeffrey
\cite{aj}, and a detailed discussion of Donaldson theory \cite{ewdon,don}
can be found in \cite[pp.~198-247]{pr}. Sections 3.2, 3.3 and 4.3 are
based on joint work with George Thompson \cite{gtlec,btmq,btqm}. Further
references can be found in the text and further information on
topological field theory in the cited reviews and the lectures of
Danny Birmingham \cite{bip} at this School.


\section{The Mathai-Quillen Formalism}

In section 2.1 we wil recall some well known facts and theorems
concerning the Euler class and the Euler number of a
finite dimensional vector bundle $E$. For our present purposes the Euler
class is most profitably understood in terms of the Thom class of $E$
and we will adopt this point of view in section 2.2. There we also
introduce and discuss at some length the \mq\ formalism which
provides, among other things, a concrete differential form realization
of the Thom class. In section 2.3 we explain how the \mq\ formalism
can be used to define certain regularized Euler numbers of $E$ when
$E$ is infinite dimensional. We will also introduce the examples
(\sqm, \tgt) which will then occupy us in the remainder of these notes.

\subsection{The Euler number of a finite dimensional vector bundle}

Consider a real vector bundle $\pi:E\ra X$ over a manifold $X$.
We will assume that $E$ and $X$ are orientable, $X$ is
compact without boundary,
and that the rank (fibre dimension) of $E$ is even and
satisfies $rk(E)=2m\leq \dim(X)=n$.

The {\em Euler class} of $E$ is an integral cohomology class
$e(E)\in H^{2m}(X,{\bf R})\equiv H^{2m}(X)$. For $m=1$ (a
two-plane bundle) $e(E)$ can e.g.~be defined in a rather
pedestrian manner (cf.~\cite{botu} for the material
covered in this and the first part of the following section). We
choose a cover of $X$ by open sets $U_{\a}$ and denote by
$g_{\a\beta}:U_{\a}\cap U_{\beta}\ra SO(2)$ the transition functions
of $E$ satisfying the cocycle condition
\be
g_{\a\beta}=g_{\beta\a}^{-1}\;\;,\;\;\;\;\;\;
             g_{\a\beta}g_{\beta\g}=g_{\a\g}\;\;.\label{1}
\ee
Identifying $SO(2)\sim U(1)$, we set $\vf_{\a\beta}=i\log g_{\a\beta}$
with
\be
\vf_{\a\beta}+\vf_{\beta\g}-\vf_{\a\g}\in 2\pi{\bf Z}\;\;,\label{2}
\ee
so that $d\vf$ is an additive cocycle,
\be
d\vf_{\a\beta}+d\vf_{\beta\g}=d\vf_{\a\g}\;\;.\label{3}
\ee
In fact, more than that is true. By introducing a partition of unity
subordinate to $\{U_{\a}\}$, i.e.~a set of functions $\rho_{\a}$
satisfying
\be
\sum_{\a}\rho_{\a}=1\;\;,\;\;\;\;\;\; supp(\rho_{\a})\ss U_{\a}\;\;,
\ee
and defining one-forms $\xi_{\a}$ on $U_{\a}$ by
$\xi_{\a}=(2\pi)^{-1}\sum_{\g}\rho_{\g}d\vf_{\g\a}$ one finds that
\be
\frac{1}{2\pi}d\vf_{\a\beta}=\xi_{\beta}-\xi_{\a}\label{5}
\ee
which obviously implies (\ref{3}). Thus $d\xi_{\a}=d\xi_{\beta}$ on the
overlaps $U_{\a}\cap U_{\beta}$ and therefore the $d\xi$'s piece together
to give a global two-form on $X$ which is closed but not necessarily
exact. The cohomology class of this form is independent of the choice of
$\xi$'s satisfying (\ref{5}) and is the Euler class
$e(E)\in H^{2}(X)$ of $E$.

For higher rank bundles a similar construction is possible in
principle but becomes rather unwieldy. Fortunately there are other, more
transparent, ways of thinking about $e(E)$.

The first of these is in terms of sections of $E$. In general, a twisted
bundle will have no nowhere-vanishing non-singular sections and one defines
the Euler class to be the homology class of the zero locus of a generic
section of $E$. Its Poincar\'e dual is then a cohomology class in
$H^{2m}(X)$.

The second makes use of the Chern-Weil theory of
curvatures and characteristic classes and produces an explicit
representative $\en(E)$ of $e(E)$ in terms of the curvature $\On$
of a connection $\nabla$ on $E$. Thinking of $\On$ as a matrix of
two-forms one has
\be
\en(E)=\frac{1}{(2\pi)^{m}}P\!f(\On)\label{6}
\ee
where $P\!f(A)$ denotes the Pfaffian of the real antisymmetric matrix $A$,
\be
P\!f(A)=\frac{(-1)^{m}}{2^{m}m!}\sum\epsilon_{a_{1}\cdots a_{2m}}
            A_{a_{1}a_{2}}\ldots A_{a_{2m-1}a_{2m}}\label{7}\;\;,
\ee
satisfying $P\!f(A)^{2} = \det(A)$. Standard arguments
show that the cohomology class of $\en$ is independent of the choice
of $\nabla$.

Finally, the third is in terms of the Thom class of $E$ and we will describe
this in section 2.2.

If the rank of $E$ is equal to the dimension of $X$ (e.g.~if $E=TX$,
the tangent bundle of $X$)  then $H^{2m}(X)=H^{n}(X)={\bf R}$ and nothing
is lost by considering, instead of $e(E)$, its evaluation on
(the fundamental class $[X]$ of) $X$, the {\em Euler number}
\be
\c(E)=e(E)[X]\label{8}\;\;.
\ee
In terms of the two descriptions of $e(E)$ given above, this number can
be obtained either as the number of zeros of a generic section $s$ of $E$
(which are now isolated) counted with multiplicity,
\be
\c(E)=\sum_{x_{k}:s(x_{k})=0}\nu_{s}(x_{k})\label{9}
\ee
(here $\nu_{s}(x_{k})$ is the degree or index of $s$ at $x_{k}$),
or as the integral
\be
\c(E)=\int_{X}\en(E)\label{10}\;\;.
\ee

Of particular interest to us is the case where $E=TX$. The Euler number
$\c(TX)$ is then equal to the {\em Euler characteristic} $\c(X)$ of $X$,
\be
\c(TX)=\c(X)\equiv\sum_{k}(-1)^{k}b_{k}(X)\label{11}
\ee
where $b_{k}(X)=\dim(H^{k}(X))$ is the $k$'th Betti number of $X$.
In this context, equations (\ref{9}) and (\ref{10}), expressing $\c(X)$
as the number of zeros of a vector field and the integral of a density
constructed from the Riemannian curvature tensor $\RX$ of $X$, are known
as the {\em Poincar\'e-Hopf theorem} and the {\em Gauss-Bonnet theorem}
respectively. For example, in two dimensions $(n=2)$, (\ref{10}) reduces
to the well known formula
\[\c(X)=\frac{1}{4\pi}\int_{X}\sqrt{g}d^{2}\!x R\]
where $R$ is the scalar curvature of $X$.

For $E=TX$ there is also an interesting generalization of (\ref{9})
involving  a vector field $V$ with a zero locus $X_{V}$ which is not
necessarily zero-dimensional. Denoting the connected components of
$X_{V}$ by $X_{V}^{(k)}$, this generalization reads
\be
\c(X)=\sum_{k}\c(X_{V}^{(k)})\label{12}\;\;.
\ee
This reduces to (\ref{9}) when the $X_{V}^{(k)}$ are isolated points
and is an identity when $V$ is the zero vector field.

One of the beauties of the \mq\ formalism, to be discussed next, is that
it provides a corresponding generalization of (\ref{10}), i.e.~an
explicit differential form representative $\ens$ of $e(E)$ depending
on both a section $s$ of $E$ and a connection $\nabla$ on $E$ such that
\be
\c(E) = \int_{X}\ens(E)\label{13}
\ee
and such that (\ref{13}) reduces to any of the above equations for
the appropriate choice of $E$ and $s$ (i.e.~to (\ref{10}) if $s$ is
the zero section, to (\ref{9}) when the zeros of $s$ are isolated,
and to (\ref{12}) for a general vectorfield on $TX$).

If $n>2m$, then we cannot evaluate $e(E)$ on $[X]$ as in (\ref{8}).
We can, however, evaluate it on homology $2m$-cycles or (equivalently)
take the product of $e(E)$ with elements of $H^{n-2m}(X)$ and evaluate
this on $[X]$. In this way one obtains {\em intersection numbers} of $X$
associated with the vector bundle $E$. A corresponding interpretation of
the Donaldson polynomials \cite{don} as observables in the topological
gauge theory of \cite{ewdon} has been given by Atiyah and Jeffrey \cite{aj}
(cf.~section 4.2).

\subsection{The Thom class and the Mathai-Quillen form}

The Euler class $e(E)$ has the property that it is the pullback of a
cohomology class on $E$, called the {\em Thom class} $\F(E)$ of $E$,
via the zero section $i:X\ra E$,
\be
e(E) = i^{*}\F(E)\;\;.\label{14}
\ee
We will show this explicitly below (cf.~equations (\ref{32},\ref{33})).
To understand the origin and significance of $\F(E)$, recall that there
are two natural notions of cohomology for differential forms on a vector
bundle $E$ over a compact manifold $X$: ordinary de Rham cohomology
$H^{*}(E)$ and compact vertical cohomology $H^{*}_{cv}(E)$. The latter
deals with forms whose restriction to any fibre has compact support.
As $E$ is contractible to $X$ one has
\be
H^{*}(E)\simeq H^{*}(X)\;\;.
\ee
On the other hand, as the compact cohomology of a vector space only has a
generator in the top dimensions (a `bump' volume form with unit volume),
one has
\be
H^{*}_{cv}(E)\simeq H^{*-2m}_{cv}(X)\;\;.\label{16}
\ee
More technically, for forms of compact vertical support one has the notion
of `push-down' or `integration along the fibres', denoted by $\pi_{*}$.
In local coordinates, and for trivial bundles, this is the obvious operation
of integrating over the fibres the part of $\omega\in\Omega^{*}_{cv}(E)$
(the space of forms with compact vertical support) which contains a vertical
$2m$-form and interpreting the result as a form on $X$. This prescription
gives a globally well defined operation
\be
\pi_{*}:\Omega^{*}_{cv}(E)\ra\Omega^{*-2m}(X)\label{17}\;\;.
\ee
In particular, for any $\omega\in\Omega^{*}_{cv}(E)$ and $\a\in\Omega^{*}(X)$
one has
\be
\pi_{*}((\pi^{*}\a)\omega) = \a\pi_{*}\omega\label{18}\;\;.
\ee
$\pi_{*}$ commutes with the exterior derivatives on $E$ and $X$
(it is sufficient to check this in local coordinates),
\be
\pi_{*}d_{E}=d_{X}\pi_{*}
\ee
and induces the so called {\em Thom isomorphism} ${\cal T}_{E}:
H^{*}(X)\ra H^{*+2m}_{cv}(E)$ (\ref{16}). Under this isomorphism,
the generator $1\in H^{0}(X)$ corresponds to a $2m$-dimensional
cohomology class on $E$, the Thom class $\F(E)$,
\be
\F(E)={\cal T}_{E}(1)\in H^{2m}_{cv}(E)\;\;.\label{20}
\ee
By definition, $\F(E)$ satisfies $\pi_{*}\F(E)=1$, so that by (\ref{18})
the Thom isomorphism is explicitly given by
\be
{\cal T}_{E}(\a) = (\pi^{*}\a)\F(E)\;\;.\label{21}
\ee

After this small digression let us now return to the Euler class $e(E)$
and equation (\ref{14}). As any two sections of $E$ are homotopic as
maps from $X$ to $E$, and as homotopic maps induce the same pullback map
in cohomology, we can use any section $s$ of $E$ instead of the zero section
to pull back $\F(E)$ to $X$ and still find
\be
s^{*}\F(E)=e(E)\label{22}\;\;.
\ee
The advantage of this way of looking at the Euler class $e(E)$ should
now be evident: provided that we can find an explicit differential form
representative $\Fn(E)$ of $\F(E)$, depending on a connection $\nabla$ on
$E$, we can pull it back to $X$ via a section $s$ to obtain a $2m$-form
\be
\ens(E)=s^{*}\Fn(E)\label{23}
\ee
representing the Euler class $e(E)$ and (if $n=2m$) satisfying (\ref{13}).
It should be borne
in mind, however, that by (\ref{22}) all these forms are cohomologous so
that this construction, as nice as it is, is not very interesting from
the cohomological point of view. To get something really new one should
therefore consider situations where the forms (\ref{23}) are not necessarily
cohomologous to $\en$. As pointed out by Atiyah and Jeffrey \cite{aj},
such a situation occurs when one considers infinite dimensional vector
bundles where $\en$ (an `infinite-form') is not defined at all. In that case
the added flexibility in the choice of $s$ becomes crucial and
opens up the pssibility of obtaining well-defined, but $s$-dependent,
`Euler classes' of $E$. We will explain this in section 2.3.

To proceed with the construction of $\Fn$, let us make two preliminary
remarks. The first is that for explicit formulae it is convenient to
switch from working with forms with compact support along the fibres
to working with `Gaussian shaped' form rapidly decreasing along the
fibres (in a suitable technical sense). Everything we have said so far
goes through in that setting \cite{mq} and we will henceforth replace
$\Omega^{*}_{cv}(E)$ by $\Omega^{*}_{rd}(E)$ etc.

The second is that Pfaffians (\ref{7}) arise as fermionic (Berezin)
integrals (this may sound like a rather mysterious remark to make at
this point, but is of course one of the reasons why what we are going
through here has anything to do with supersymmetry and topological field
theory). More precisely, if we have a real antisymmetric matrix $(A_{ab})$
and introduce real Grassmann odd variables $\c^{a}$, then
\be
P\!f(A)=\int d\c e^{\c^{a}A_{ab}\c^{b}/2}\label{24}\;\;.
\ee
In particular, we can therefore write the form $\en$ (\ref{6}) as
\be
\en(E)=(2\pi)^{-m}\int d\c e^{\c_{a}\On^{ab}\c_{b}/2}\label{25}.
\ee
The idea is now to extend the right hand side of (\ref{25}) to a
form $\Fn(E)$ on $E$ having Gaussian decay along the fibres and
satisfying $\pi_{*}\Fn(E)=1$.

Regarding $E$ as a vector bundle associated to a principal $G$ bundle
$P$ with standard fibre $F$, $E=P\times_{G}F$, we can represent forms
on $E$ by basic, i.e.~horizontal and $G$-invariant, forms on $P\times F$,
\be
\Omega^{*}(E)=\Omega^{*}_{bas}(P\times F)\label{26}
\ee
and sections of $E$ by $G$-equivariant maps from $P$ to $F$.
Moreover, via the projection $\pi:P\ra X$, $E$ pulls back to the
canonically trivial vector bundle $\pi^{*}E = P\times F$ over $P$
whose induced connection and curvature we also denote by $\nabla$ and
$\On$. With this identification understood, the Thom form $\Fn(E)$
of Mathai and Quillen is given by
\be
\Fn(E)=(2\pi)^{-m}e^{-\xi^{2}/2}\int d\c
       e^{\c_{a}\On^{ab}\c_{b}/2 +i\nabla\xi^{a}\c_{a}}\label{27}
\ee
where we have chosen a fixed fibre metric on $F$,
$\xi^{a}$ are coordinates on $F$ and $\nabla\xi^{a}$ is the
exterior covariant derivative of $\xi^{a}$, a one-form on $P\times F$.
We now check that $\Fn(E)$ really represents the Thom class of $E$.

First of all, integrating out $\c$ one sees that (\ref{27}) defines a
$2m$-form on $P\times F$. This form is indeed basic and represents
a closed $2m$-form on $E$. $G$-invariance and horizontality are
almost obvious from (\ref{27}) as $\On$ and $\nabla\xi$ are horizontal
(by the definition of the covariant exterior derivative). Less evident
is the fact that $\Fn(E)$ is closed. This is best understood in
terms of the equivariant cohomology $H_{G}^{*}(F)$ of $F$ (cf.~sections
5 and 6 of \cite{mq}) and is related to the fact that the exponent
in (\ref{27}),
\be
-\xi^{2}/2+\c_{a}\On^{ab}\c_{b}/2 +i\nabla\xi^{a}\c_{a}\;\;,\label{28}
\ee
is invariant under the graded (i.e.~super-) symmetry
\bea
\d\c_{a}&=&i\xi_{a}\nonumber\\
\d\xi^{a}&=&\nabla\xi^{a}\label{29}
\eea
mapping the Grassmann odd $\c$ to the even $\xi$ and $\xi$ to the
Grassmann odd one-form $\nabla\xi$. `On shell', i.e.~using the
$\c$ equation of motion $i\nabla\xi^{a}=\On^{ab}\c_{b}$, this
supersymmetry squares to rotations by the curvature matrix $\On$,
\bea
\d^{2}\c^{a}&=&\On^{ab}\c_{b}\nonumber\\
\d^{2}\xi^{a}&=&\On^{ab}\xi_{b}\label{30}
\eea
which is the hallmark of equivariant cohomology. For a more thorough
discussion of the relation between the classical (Cartan-, Weil-)
models of equivariant cohomology and the BRST model, as well as of
the Mathai-Quillen formalism in that context, see \cite{jk}.

By introducing a
Grassmann even scalar field $B_{a}$ with $\d\c_{a}=B_{a}$ and
$\d B^{a} = \On^{ab}\c_{b}$ the `action' (\ref{28}) becomes
$\d$-exact off-shell,
\be
(\ref{28})\approx \d (\c_{a}(i\xi_{a}-B^{a}/2)\label{exact}
\ee
It is of course
no coincidence that the structure we have uncovered here is reminiscent
of \tft, see e.g.~(\ref{54},\ref{78}) below.

Because of the factor $e^{-\xi^{2}/2}$, (\ref{27}) is certainly rapidly
decreasing along the fibre directions. What remains to be checked
to be able
to assert that $\Fn(E)$ represents the Thom class $\F(E)$ is
that $\pi_{*}\Fn(E)=1$ or, under the isomorphism (\ref{26}),
that $\int_{F}\Fn(E)=1$. Extracting from the $2m$-form $\Fn(E)$ the part
which is a $2m$-form on $F$ we find that indeed
\bea
\int_{F}\Fn(E)&=&(2\pi)^{-m}\int_{F}e^{-\xi^{2}/2}\int d\c
  \frac{(id\xi^{a}\c_{a})^{2m}}{2m!}\nonumber\\
  &=&(2\pi)^{-m}\int_{F}e^{-\xi^{2}/2}d\xi^{1}\ldots d\xi^{2m} = 1\;\;.
\eea
This proves that
\be
[\Fn(E)]=\F(E)\in H^{2m}_{rd}(E)\label{32}\;\;.
\ee

We now take a closer look at the forms $s^{*}\Fn(E)=\ens(E)$ (\ref{23})
for various choices of $s$. In our notation $\ens(E)$ is obtained
from (\ref{27}) by replacing the fibre coordinate $\xi$ by $s(x)$.
The first thing to note is that for the zero section $i$, (\ref{27})
reduces to (\ref{25}) and therefore
\be
\en(E)=i^{*}\Fn(E)\label{33}\;\;.
\ee
This is a refinement of (\ref{14}) to an equality between differential
forms and therefore, in particular, finally proves (\ref{14}) itself.

If $n=2m$ and $s$ is a generic section of $E$ transversal to the
zero section, then we can calculate $\int_{X}\ens(E)$ by replacing
$s$ by $\g s$ for $\g\in{\bf R}$ and evaluating the integral in the limit
$\g\ra\infty$. In that limit the curvature term in (\ref{27}) will
not contribute and one can use the stationary phase approximation
to reduce the integral to a sum of contributions from the zeros of $s$,
reproducing equation (\ref{9}). The calculation is entirely analogous
to similar calculations in \sqm\ (see e.g.~\cite{pr}) and I will not
repeat it here. In fact, as we will later derive the Mathai-Quillen
formula (\ref{27}) from \sqm\ (section 3.3), this shows that
the required manipulations
are not only entirely analogous to but identical with those in \sqm.
As we could equally well have put $\g =0$ in the above, this also
establishes directly the equality of (\ref{9}) and (\ref{10}).

Finally, if $E=TX$ and $V$ is a non-generic section of $X$ with zero
locus $X_{V}$, the situation is a little bit more complicated. It
turns out that in this case $\int_{X}e_{V,\nabla}$ can be expressed in
terms of the Riemann curvature tensor ${\cal R}_{X_{V}}$ of $X_{V}$.
Here ${\cal R}_{X_{V}}$ arises from the data $\RX$ and $V$ entering
$e_{V,\nabla}$ via the {\em Gauss-Codazzi equations}. Quite generally,
these express the curvature $\RY$ of a submanifold $Y\ss X$ in terms
of $\RX$ and the extrinsic curvature of $Y$ in $X$ (we will recall
these equations in section 4.1). Then equation (\ref{12}) is reproduced
in the present setting in the form (we assume that $X_{V}$ is connected
- this is for notational simplicity only)
\be
\c(X)=\int_{X}e_{V,\nabla}=(2\pi)^{-\dim(X_{V})/2}\int_{X_{V}}
P\!f({\cal R}_{X_{V}})\label{34}\;\;.
\ee
Again the manipulations required to arrive at (\ref{34}) are exactly
as in \sqm\ \cite{btmq,btqm} and we will perform such a calculation in
the context of topological gauge theory in section 4.3 (see the calculations
leading to (\ref{89})).

\subsection{The Mathai-Quillen formalism for infinite dimensional vector
bundles}

Let us recapitulate briefly what we have achieved so far. Using the
\mq\ form $\Fn(E)$ (\ref{27}), we have constructed a family of
differential forms $\ens(E)$ parametrized by a section $s$ and a connection
$\nabla$ and all representing the Euler class $e(E)\in H^{2m}(X)$. In
particular, for $E=TX$, the equation $\c(X)=\int_{X}\env(X)$ interpolates
between the classical Poincar\'e-Hopf and Gauss-Bonnet theorems.

To be in a situation where the forms $\ens$ are not necessarily all
cohomologous to $\en$, and where the \mq\ formalism thus `comes into its
own' \cite{aj}, we now consider infinite dimensional vector bundles. To
motivate the concept of regularized Euler number of such a bundle,
to be introduced below, recall equation (\ref{12}) for the Euler number
$\c(X)$ of a manifold $X$ which we repeat here for convenience in the form
\be
\c(X)=\c(X_{V})\label{35}\;\;.
\ee
When $X$ is finite dimensional this is an identity, while its left hand
side is not defined when $X$ is infinite dimensional. Assume, however,
that we can find a vector field $V$ on $X$ whose zero locus is a finite
dimensional submanifold of $X$. Then the right hand side of (\ref{35})
{\em is} well defined and we can use it to tentatively define a
{\em regularized Euler number} $\c_{V}(X)$ as
\be
\c_{V}(X):=\c(X_{V})\label{36}\;\;.
\ee
By (\ref{13}) and
the standard localization arguments, as reflected e.g.~in (\ref{34}),
we expect this number to be given by the (functional) integral
\be
\c_{V}(X)=\int_{X}\env(X)\label{37}\;\;.
\ee
This equation can (formally) be confirmed by explicit calculation.
The idea is again to replace $V$ by $\g V$, so that (\ref{37})
localizes to the zeros of $V$ as $\g\ra\infty$, and to show that in this
limit the surviving terms in (\ref{28}) give rise to the Riemann curvature
tensor of $X_{V}$, expressed in terms of $\RX$ and $V$ via the Gauss-Codazzi
equations.
A rigorous proof can probably be obtained in some
cases by probabilistic methods as used e.g.~by Bismut \cite{bismut1,bismut2}
in related contexts. We will, however,
content ourselves with verifying (\ref{37})
in some examples below.

More generally, we are now led to define the regularized Euler number
$\cs(E)$ of an infinite dimensional vector bundle $E$ as
\be
\cs(E):=\int_{X}\ens(E)\;\;.\label{38}
\ee
Again, this expression turns out to make sense when the zero locus of
$s$ is a finite dimensional manifold $X_{s}$, in which case $\cs(E)$ is
the Euler number of some finite dimensional vector bundle over $X_{s}$
(a quotient bundle of the restriction $E|_{X_{s}}$, cf.~\cite{ewwzw,aspin}).

Of course, there is no reason to expect $\cs(E)$ to be independent of $s$,
even if one restricts one's attention to those sections $s$ for which the
integral (\ref{38}) exists. However, if $s$ is a section of $E$ naturally
associated with $E$ (we will see examples of this below), then $\cs(E)$
is also naturally associated with $E$ and can be expected to
carry interesting topological information. This is indeed the case.

It is precisely such a representation of finite dimensional topological
invariants by infinite dimensional integrals which is the characteristic
property of topological field theories. It is then perhaps not too
surprising anymore at this point, that topological field theory actions
can be constructed from (\ref{28}) for suitable choices of $X$, $E$, and $s$.

Here is a survey of the examples we will discuss in a little more detail
in the following sections ($LM$ denotes the loop space of a manifold $M$
and $\A{k}$ a space of gauge orbits in $k$ dimensions).

\noindent{\bf Example 1} \underline{$X=LM$, $E=TX$, $V=\dx$}
(section 3.2)\\
(\ref{28}) becomes the standard action $S_{M}$ of de Rham \sqm\ and
\be
\int_{LM}\env(LM) = Z(S_{M})\label{39}
\ee
is the partition function of $S_{M}$. The zero locus $(LM)_{V}$ of $V$ is
the space of constant loops, i.e.~$(LM)_{V}\simeq M$. We therefore expect
(\ref{39}) to calculate
\be
\c_{V}(LM)=\c(M)\label{40}\;\;.
\ee
As this indeed agrees with the well known explicit evaluation of
$Z(S_{M})$ in the form
\be
Z(S_{M})=(2\pi)^{-\dim(M)/2}\int_{M}P\!f({\cal R}_{M})\label{41}\;\;,
\ee
this is our first confirmation of (\ref{38}). Conversely the \mq\ formalism
now provides an understanding and explanation of the mechanism by which
the (path) integral (\ref{39}) over $LM$ localizes to the integral (\ref{41})
over $M$.

Instead of the vector field $\dx$ one can also use $\dx + W'$, where
$W'$ denotes the gradient vectorfield of some function $W$ on $M$. By
an argument to be introduced in section 3 (the `squaring argument') the
zero locus of this vector field is the zero locus of $W'$ on $M$
(i.e.~$\dx = W' =0$) whose Euler number is the same as that of $M$ by
(\ref{35}),
\be
\c_{V}(LM)=\c(M_{W'})=\c(M)\;\;.\label{42}
\ee
Again this agrees with the explicit evaluation of the path integral
of the corresponding \sqm\ action.

\noindent{\bf Example 2}
\underline{$X=\A{4}$, $E={\cal E}_{+}$, $s=(F_{A})_{+}$} (section 4.2)\\
(${\cal E}_{+}$ is a certain bundle of self-dual two-forms over $\A{4}$
and $(F_{A})_{+}$ is the self-dual part of the curvature $F_{A}$ of $A$).
The zero locus $X_{s}$
is the moduli space $\M_{I}$ of instantons, and not unexpectedly the
corresponding action is that of Donaldson theory \cite{don,ewdon}.
The partition function $\cs({\cal E}_{+})$ is what is known as the first
Donaldson invariant and is only non-zero when $d(\M)\equiv \dim(\M_{I})=0$.
If $d(\M)\neq 0$ then one has to insert elements of
$H^{d(\M)}(\A{4})$ into the path integral in the manner explained at
the end of section 2.1 to obtain non-vanishing results (the Donaldson
polynomials). This interpretation of Donaldson theory is due to Atiyah and
Jeffrey \cite{aj}.

\noindent{\bf Example 3} \underline{$X=\A{3}$, $E=TX$, $V=*F_{A}$}
(section 4.3)\\
($*$ is the Hodge operator, and the one-form $*F_{A}$ defines a vector
field on $\A{3}$, the gradient vector field of the Chern-Simons functional).
The zero locus of $V$ is the moduli space $\M^{3}$ of flat
connections and the action coincides with that constructed in
\cite{ewtop,bbt,btmq}. Again one finds full agreement of
\be
\c_{V}(\A{3}) = \c(\M^{3})\label{43}
\ee
with the partition function of the action which gives $\c(\M^{3})$ in the
form (\ref{34}), i.e.~via the Gauss-Codazzi equations for the embedding
$\M^{3}\ss\A{3}$. In \cite{aj} this partition function was first
identified with a regularized Euler number of $\A{3}$. We have now
identified it more specifically with the Euler number of $\M^{3}$.
In \cite{tau} it was shown that for certain
three-manifolds (homology spheres) $\c_{V}(\A{3})$ is the Casson invariant.
Hence our considerations suggest that the Casson invariant can be defined as
$\c(\M^{3})$ for more general three-manifolds \cite{btmq}.

\noindent{\bf Example 4}
\underline{$X=L(\A{3})$, $E=TX$, $V=\dot{A}+*F_{A}$} (section 4.3)\\
This is \sqm\ on $\A{3}$ and in a sense a combination of all the three
above examples. The resulting (non-covariant) gauge theory action in
$3+1$ dimensions is that of Donaldson theory (example 2).
After partial localization from $L(\A{3})$ to $\A{3}$ it is seen to be
equivalent to the action of example 3. Further reduction to the zeros
of the gradient vector field $*F_{A}$ (example 1)
reduces the partition function to an integral over $\M^{3}$ and calculates
$\c(\M^{3})$. This again confirms the equality of the left and right hand
sides of (\ref{37}). The reason why Donaldson theory is related to instanton
moduli spaces in example 2, but to moduli spaces of flat connections in
this example is explained in \cite{btmq}.

\section{The Euler Number of Loop Space and Supersymmetric
Quantum Mechanics}

In this section we will work out some of the details of example 1. We begin
with a (very) brief survey of the geometry of loop space (section 3.1).
We then apply the \mq\ formalism to the tangent bundle of loop space, derive
\sqm\ from that, and review some of the most important features of \sqm\
in the light of this derivation (section 3.2). Finally, to complete the
picture, we explain how the finite-dimensional \mq\ form (\ref{27})
can be derived from \sqm\ (section 3.3).

\subsection{Loop space geometry}

We denote by $M$ a smooth orientable Riemannian manifold with metric
$g$ and by $LM$
the loop space of $M$, i.e.~the space of smooth maps from the circle
$S^{1}$ to $M$,
\be
LM:=C^{\infty}(S^{1},M)\label{44}
\ee
(consistent with the sloppyness to be encountered throughout these notes
we will not worry about the technicalities of infinite dimensional
manifolds). Elements of $LM$ are denoted by $x(t)$ or $x^{\m}(t)$,
where $t\in [0,1]$, $x^{\m}$ are (local) coordinates on $M$ and
$x^{\m}(0)=x^{\m}(1)$. In \sqm\ it is
convenient to scale $t$ such that $t\in [0,\beta]$ and
$x^{\m}(0)=x^{\m}(\beta)$ for some $\beta\in{\bf R}$, and to regard
$\beta$ as an additional parameter (the inverse temperature) of the theory.

A tangent vector to a loop $x(t)$ can be regarded as an infinitesimal
variation of the loop. As such it can be thought of as a
vector field on the image $x(\s1)\ss M$ (tangent to $M$ but not
necessarily to the loop $x(\s1)$). In other words, the tangent space
$T_{x}(LM)$ to $LM$ at the loop $x(t)$ is the space of smooth sections of
the tangent bundle $TM$ restricted to the loop $x(t)$,
\be
T_{x}(LM)\simeq\Gamma^{\infty}(x^{*}(TM))\label{45}\;\;.
\ee
There is a canonical vector field on $LM$ which generates rigid rotations
$x(t)\ra x(t+\epsilon)$ of the loop around itself. It is given by
$V(x)(t)=\dx(t)$ (or $V=\dx$ for short).
The metric $g$ on $M$ induces a metric $\hat{g}$ on $LM$ through
\be
\hat{g}_{x}(V_{1},V_{2}) = \tint{1} g_{\m\n}(x(t))
            V_{1}^{\m}(x)(t)V_{2}^{\n}(x)(t)
\label{46}\;\;.
\ee
Likewise, every $p$-form $\a$ on $M$ gives rise to a $p$-form $\hat\a$ on $LM$
via
\be
\hat\a_{x}(V_{1},\ldots,V_{p})=\tint{1}
\a_{x(t)}(V_{1}(x)(t),\ldots,V_{p}(x)(t))
\label{47}\;\;,
\ee
and a local basis of one-forms on $LM$ is given by the differentials
$dx^{\m}(t)$.

The last piece of information we need is that the
Levi-Civit\`a connection on $M$ can be pulled back to
$\s1$ via a loop $x(t)$. This defines a covariant derivative on (\ref{45})
and its dual which we denote by $\nt$. We have e.g.
\be
(\nt V^{\m})(x)(t)={\textstyle\frac{d}{dt}}V^{\m}(x)(t) +
\Gamma^{\m}_{\n\rho}(x(t))\dx^{\n}(t)V^{\rho}(x)(t)\label{48}\;\;.
\ee

\subsection{Supersymmetric quantum mechanics}

We are now in a position to discuss example 1 of section 2.3 in more detail.
In the notation of that section, we choose $X=LM$, $E=TX$, and $V=\dx$. The
anticommuting variables $\c_{a}$ thus parametrize the fibres of $TX$ and we
write them as $\c_{a}=e_{a}^{\;\m}\pb_{\m}$ where $e_{a}^{\;\m}$ is the
inverse vielbein corresponding to $g_{\m\n}$. Using the metric (\ref{46})
as a fibre metric on $T_{x}X$, the first term of (\ref{28}) is simply the
standard bosonic kinetic term of quantum mechanics ,
\be
\xi^{2}/2\ra\tint{\beta}g_{\m\n}\dx^{\m}\dx^{\n}/2
\label{49}\;\;.
\ee
To put the remaining terms into a more familiar form, we use the standard
trick of replacing the differentials $dx^{\m}(t)$ by periodic anticommuting
variables,
\be
dx^{\m}(t)\ra\p^{\m}(t)\label{50}
\ee
and integrating over them as well. As the integral over the $\p$'s will
simply pick out the top-form part which is then to be integrated over $X$
(cf.~(\ref{38})), nothing is changed by the substitution (\ref{50}). With
all this in mind the complete exponential of the \mq\ form $\env(LM)$
becomes
\be
S_{M}=\tint{\beta}[-g_{\m\n}\dx^{\m}\dx^{\n}/2
            +R^{\m\n}_{\;\;\rho\sigma}\pb_{\m}\p^{\rho}\pb_{\n}\p^{\sigma}/4
            -i\pb_{\m}\nt\p^{\m}]\label{51}\;\;.
\ee
This is precisely the standard action of de Rham (or $N=1$) \sqm\, to be
found e.g.~in \cite{ewqm,ewqm1,ag1,ag2,fw} (with the spinors appearing there
decomposed into their components; we also choose $\p$ and $\pb$ to be
independent real fields instead of complex conjugates). It will be
convenient to introduce a multiplier field $B_{\m}$ and to rewrite the action
(\ref{51}) in first order form,
\be
S_{M} =  \tint{\beta}[ i \dx^{\m}B_{\m} +
g^{\m\n}B_{\m}B_{\n}/2+
R^{\m\n}_{\;\;\rho\sigma}\bar{\psi}_{\m}\psi^{\rho}\bar{\psi}_{\n}
\psi^{\sigma}/4  -i \bar{\psi}_{\m}\nabla_{t}\psi^{\m} ] \;\;.
\label{52}
\ee
The supersymmetry of this action is
\bea
\d x^{\m} &=& \p^{\m}\;,\;\;\;
\d\pb_{\m}=B_{\m}-\Gamma^{\n}_{\;\m\rho}\pb_{\n}\p^{\rho}\nonumber\\
\d \p^{\m}&=& 0\;\;,\;\;\;\;
\d B_{\m}=\Gamma^{\n}_{\;\m\rho}B_{\n}\p^{\rho}
            -R^{\n}_{\;\m\rho\sigma}\pb_{\n}\p^{\rho}\p^{\sigma}/2
\label{53}\;\;.
\eea
This is readily verified by noticing that $\d^{2}=0$ and that (\ref{52})
can itself be written as a supersymmetry variation,
\be
S_{M}=\d\tint{\beta}[\pb_{\m}(i\dx^{\m}+g^{\m\n}B_{\n}/2)]\label{54}
\;\;.
\ee
Note the similarity with (\ref{exact}). Reinterpreting
$\d$ as a BRST operator, this also shows that the sector of
\sqm\ annihilated by $\d$ is topological (a BRST exact action being one of the
hallmarks of \tft).  As we will see below that only groundstates contribute
to the partition function anyway, it is, in particular, independent of the
coefficient of the second term of (\ref{54}) regardless of whether
we treat $\d$ as a conventional supersymmetry operator (mapping bosonic to
fermionic states and vice-versa) or as a BRST operator (annihilating
physical states). Rescaling this term by a
real parameter $\a$ we find the equivalent action
\be
S_{M}=\tint{\beta}[-g_{\m\n}\dx^{\m}\dx^{\n}/2\a
          +\a R^{\m\n}_{\;\;\rho\sigma}\pb_{\m}\p^{\rho}\pb_{\n}\p^{\sigma}/4
          -i\pb_{\m}\nt\p^{\m}]\label{55}\;\;.
\ee
On the other hand, if we rescale the time variable by $\beta$ we obtain
the action (\ref{55}) with $\tint{\beta}$ replaced by $\tint{1}$ and $\a$
replaced by $\beta$. Thus the `topological' $\a$-independence translates
into the quantum mechanical $\beta$-independence. Conversely, this
$\beta$-independence is obvious from the standard Hamiltonian construction
of \sqm\ (cf.~below) and translates into
the topological $\a$-independence of (\ref{55}).

This is not the place to enter into a detailed discussion of \sqm, and
we will in the following focus on those aspects relevant for the
\mq\ side of the issue and
our subsequent considerations involving topological gauge theories.
For detailed discussions of \sqm\ in the context of index theory and
topological field theory the reader is referred to \cite{ag2} and
\cite[pp.~140-176]{pr} respectively.

Our discussion of the \mq\ formalism suggests that the partition function
$Z(S_{M})$ of the \sqm\ action $S_{M}$ (\ref{51}), with periodic boundary
conditions on all the fields, is the Euler number $\c(M)$
of $M$ (as $\c_{V}(LM)=\c((LM)_{V})=\c(M)$, cf.~(\ref{36}-\ref{40})).
As is well known, this is indeed the case.

The conventional way to see this (if one does not yet trust the
infinite dimensional version of the \mq\ formalism) is to start with the
definition of $\c(M)$ as the Euler characteristic of
$M$ (\ref{11}). As there is a one-to-one
correspondence between cohomology classes and harmonic forms on $M$
(more precisely, there is a unique representative in every
de Rham cohomology class which is annihilated by the Laplacian
$\Delta = d d^{*}+d^{*}d$) one can write $\c(X)$ as a trace
over the space ${\rm Ker}\;\Delta$,
\be
\c(M)={\rm tr}_{{\rm Ker}\;\Delta}(-1)^{F}\label{56}\;\;,
\ee
where $(-1)^{F}$ is $+1$ ($-1$) on even (odd) forms. As the operator
$d+d^{*}$ commutes with $\Delta$ and maps even to odd forms and vice-versa,
there is an exact pairing between `bosonic' and `fermionic' eigenvectors
of $\Delta$ with non-zero eigenvalue. It is thus possible to extend the
trace in (\ref{56}) to a trace over the space of all differential forms,
\be
\c(M)={\rm tr}_{\Omega^{*}}(-1)^{F}e^{-\beta\Delta}\label{57}\;\;.
\ee
As only the zero modes of $\Delta$ will contribute to the trace, it
is evidently independent of the value of $\beta$. Once one has put
$\c(M)$ into this form of a statistical mechanics partition function,
one can use the Feynman-Kac formula to represent it as a supersymmetric
path integral \cite{ag2} with the action (\ref{51}), imaginary time
of period $\beta$ and periodic boundary conditions on the anticommuting
variables $\p^{\m}$ (due to the insertion of $(-1)^{F}$). Conversely,
a Hamiltonian analysis of the action (\ref{51}) would tell us that
we can represent its Hamiltonian by the Laplacian $\Delta$ on differential
forms \cite{ewqm1} and, tracing back the steps which led us to (\ref{57}),
we would then again deduce that $Z(S_{M})=\c(M)$, as anticipated in
(\ref{39}).

This Hamiltonian way of arriving at the action of \sqm\ should be
contrasted with the \mq\ approach. In the former
one starts with the operator whose index one
wishes to calculate (e.g.~$d+d^{*}$), constructs a corresponding Hamiltonian,
and then deduces the action. On the other hand,
in the latter one begins with a finite dimensional
topological invariant (e.g.~$\c(M)$) and represents that directly
as an infinite
dimensional integral, the partition function of a supersymmetric action.

What makes such a path integral representation of $\c(M)$ interesting is
that one can now go ahead and try to somehow evaluate it
directly, thus possibly obtaining alternative expressions for $\c(M)$.
Indeed, one can obtain path integral `proofs' of the Gauss-Bonnet and
Poincar\'e-Hopf theorems in this way. This is just the infinite dimensional
analogue of the considerations of section 2.2 where different choices of $s$
in $\int_{X}\ens(E)$ lead to different expressions for $\c(E)$. As we will
derive the finite dimensional \mq\ form from \sqm\ in section 3.3 we can
appeal to the manipulations of section 2.2 to complete these `proofs'.
However, it is also instructive to perform these calculations directly.
Before indicating how this can be done, we will need to introduce a
generalization of the action (\ref{51}) which arises when one takes the
section $\dx^{\m}+\gamma g^{\m\n}\del_{\n}W$
of $T(LM)$ (cf.~example 1 of section 2.3) to regularize the Euler number
of $LM$. Here $W$ is a function (potential) on $M$ and $\gamma$ is yet one
more arbitrary real parameter. In that case one obtains (introducing also the
parameter $\a$ of (\ref{55}))
\bea
S_{M,\gamma W}& = & \tint{\beta}[i(\dx^{\m}+
                   \gamma g^{\m\n}\del_{\n}W(x))B_{\m} +
                \a g^{\m\n}B_{\m}B_{\n}/2+\a
R^{\m\n}_{\;\;\rho\sigma}\pb_{\m}\psi^{\rho}\pb_{\n}\psi^{\sigma}/4
\nonumber\\
& & -i \bar{\psi}_{\m}(\d^{\m}_{\n}\nabla_{t}+\gamma g^{\m\rho}
\nabla_{\rho}\del_{\n}W)\psi^{\n} ] \;\;.
\label{58}
\eea
{}From the Hamiltonian point of view this action arises from replacing the
exterior derivative $d$ by
\be
d\ra d_{\gamma W}\equiv e^{-\gamma W}de^{\gamma W}\label{59}\;\;.
\ee
and applying the above procedure to the corresponding Laplacian
$\Delta_{\gamma W}$.
As there is a one-to-one correspondence between $\Delta$- and
$\Delta_{\gamma W}$-harmonic forms, this also represents $\c(M)$
(independently of the value of $\gamma$).

This freedom in the choice of parameters $\a,\beta ,\g$ greatly facilitates
the evaluation of the partition function. Let us, for example, choose
$\a=0$ in (\ref{58}). Then the curvature term drops out completely and the
$B$-integral will simply give us a delta function
constraint $\dx^{\m}+\g g^{\m\n}\del_{\n}W=0$. Squaring this equation and
integrating it over $t$ one finds
\bea
&&\dx^{\m}+\g g^{\m\n}\del_{\n}W=0\nonumber\\
&\ra&\tint{\beta}g_{\m\n}\dx^{\m}\dx^{\n}+\g^{2}g^{\m\n}\del_{\m}W\del_{\n}W
             + 2\g \dx^{\m}\del_{\m}W=0\nonumber\\
&\ra&\dx^{\m}=0=\del_{\m}W\label{60}
\eea
as the second line is the sum of two nonnegative terms and a total
derivative. This is the `squaring argument' referred to in section 2.3. It
demonstrates that the path integral over $LM$ is reduced to an integral
over $M$ (by $\dx^{\m}=0$) and further to an integral over the set $M_{W'}$
of critical points of $W$ (and analogously for the $\p$'s by
supersymmetry). When the critical points are isolated,
inspection of (\ref{58}) immediately reveals that the partition function is
\be
\c(M)=Z(S_{M,W})=\sum_{x_{k}:dW(x_{k}=0}{\rm sign}(\det H_{x_{k}}(W))
\label{61}\;\;,
\ee
where
\be
H_{x_{k}}(W)=(\nabla_{\m}\del_{\n}W)(x_{k})\label{62}
\ee
is the {\em Hessian} of $W$ at $x_{k}$. This is the Poincar\'e-Hopf theorem
(\ref{9}). This result can also be derived by keeping $\a$ non-zero and
taking the limit $\g\ra\infty$ instead which also has the effect of
localizing the path integral around the critical points of $W$ because
of the term $\g^{2}W'^{2}$ in the action.

If we switch off the potential, then we can not simply set $\a=0$ in
(\ref{58}), as the resulting path integral would be singular due
to the undamped bosonic and fermionic zero modes. In that case, the limit
$\a\ra 0$ or $\beta\ra 0$ has to be taken with more care. Since whatever
we can do with $\a$ we can also do with $\beta$ let us set $\a=1$ in the
following. We first rescale the time coordinate $t$ by $\beta$, and then
we rescale $B$ and $\pb$ by $\beta^{1/2}$, $B\ra\beta^{1/2}B$ and
$\pb\ra\beta^{1/2}\pb$, and all the non-zero-modes of $x$ and $\p$ by
$\beta^{-1/2}$. This will leave the path integral measure invariant
and has the effect that all the $\beta$-dependent terms in the action
are at least of order $O(\beta^{1/2})$ and the limit $\beta\ra 0$ can now
be taken with impunity. The integral over the non-constant modes gives 1 and
the net-effect of this is that one is left with a finite-dimensional
integral of the form (\ref{25}), namely
\be
\c(M)\sim\int\! dx\int\! d\p \int\! d\pb\; e^{
R^{\m\n}_{\;\;\rho\sigma}\pb_{\m}\psi^{\rho}\pb_{\n}\psi^{\sigma}/4}
\;\;,\label{fred}
\ee
over the constant modes of $x$, $\p$, and $\pb$ of which there are
$\dim(M)$ each. In order to get a non-zero contribution (i.e. to soak up
the fermionic zero modes) one has to expand (\ref{fred}) to $(\dim(M)/2)$'th
order, yielding the Pfaffian of ${\cal R}_{M}$ and hence, upon integration
over $M$ (the $x$ zero modes) the Gauss-Bonnet theorem (\ref{6},\ref{10}).
(\ref{fred}) also gives the correct result for odd dimensional manifolds,
$\c(M)=0$, as there is no way to pull down an odd number of $\p$'s and
$\pb$'s from the exponent.

If the critical points of $W$ are not isolated then, by a combination of the
above arguments, one recovers the generalization
$\c(M)=\c(M_{W'})$ (\ref{12},\ref{42}) of the
Poincar\'e-Hopf theorem in the form (\ref{34}).

As this treatment of \sqm\ has admittedly been somewhat sketchy I should
perhaps, summarizing this section, state clearly what are the
important points to keep in mind:
\begin{enumerate}
\item The \mq\ formalism applied to the loop space $LM$ of a Riemannian
manifold $M$ leads directly to the action of \sqm\ with target space $M$.
Different sections lead to different actions, and those we have considered
all regularize the Euler number of $LM$ to be $\c(M)$.
\item Explicit evaluation of the \sqm\ path integrals obtained in this way
confirms that we can indeed represent the regularized Euler number
$\c_{V}(LM)$, as defined by (\ref{36}), by the functional integral (\ref{37}).
\item Finally, I have argued (although not proved in detail) that the zero
modes are all that matter in \sqm, the integral over the non-zero-modes
giving $1$. This observation is useful when one attempts to construct
topological gauge theories from \sqm\ on spaces of connections
(see \cite{btqm} and the remarks in section 4.3).
\end{enumerate}

\subsection{The Mathai-Quillen form from \sqm}

So far we have derived the action of supersymmetric quantum mechanics by
formally applying the Mathai-Quillen formalism to $LM$, and we have
indicated how to rederive the classical (generalized) Poincar\'e-Hopf
and Gauss-Bonnet formulae. What is still lacking to complete the picture
is a derivation of the general (finite dimensional) Mathai-Quillen form
$\Fn(E)$ (\ref{27}) for $E=TM$ from \sqm.

As $\Fn(TM)$ can be pulled back to $M$ via an arbitrary vector field
(section of $TM$) $v$, not necessarily a gradient vector field,
we need to consider the \sqm\ action resulting from the regularizing
section  $\dx + \g v$ of $T(LM)$. This is just the action (\ref{58})
with $\del_{\n}W$ replaced by $g_{\m\n}v^{\m}$.
In that case the squaring argument, as expressed in (\ref{60}),
fails because the cross-term will not integrate to zero. In the
limit $\g\ra\infty$ the path integral will nevertheless reduce to a
Gaussian around the zero locus of $v$ because of the term
$\g^{2} g_{\m\n}v^{\m}v^{\n}$ in the action, and in this limit the path
integral calculates $\c(M)=\c(M_{v})$ in the form (\ref{34}).

To derive the \mq\ form, however, we are interested in finite values of
$\g$. Thus, what we need to do now
is adjust the parameters in such a way that the zero modes of
all the terms involving the vector field $v$ or the curvature survive.
Proceeding exactly as in the derivation of the Gauss-Bonnet theorem
one ends up with a time-independent `action' of the form
\be
B^{2}/2+\g v^{\m}B_{\m}
+ R^{\m\n}_{\;\;\rho\sigma}\pb_{\m}\pb_{\n}\p^{\rho}\p^{\sigma}/4
-i \g\pb_{\m}\nabla_{\n}v^{\m}\p^{\n} \label{63}
\ee
which - upon integration over $B$ - reproduces
precisely the exponent (\ref{28}) of the Mathai-Quillen form (\ref{27})
with $\xi^{a}$ replaced by the arbitrary section $\g v^{\m}$ or
$\g e^{a}_{\;\m}v^{\m}$ of $TM$. We have thus also rederived the \mq\
formula (\ref{13}) for $TM$,
\be
\c(M)=\int_{M}e_{v,\nabla}(TM)\;\;,\label{64}
\ee
from \sqm. Specializing now to $v=0$ or $v$ a generic vector field with
isolated zeros again reproduces the classical expressions.

\section{The Euler Number of Vector Bundles over $\C$ and
Topological Gauge Theory}

In this section we essentially work out the details of examples
2 and 3 and discuss some related models as well.
Section 4.1 contains a brief summary of the facts we will need
from the geometry of gauge theories . In section 4.2 we will see
how Donaldson theory can be interpreted in terms of the \mq\ formalism.
Section 4.3 sketches the construction of a topological
gauge theory in $3d$ from the tangent bundle over or (alternatively)
\sqm\ on gauge orbit space which represents the Euler characteristic of
the moduli space of flat connections. It also contains a brief discussion of
the $2d$ analogue of Donaldson theory.

\subsection{Geometry of gauge theories}

Let $(M,g)$ be a compact, oriented, Riemannian
manifold, $\pi:P\ra M$ a principal $G$ bundle over $M$, $G$ a compact
semisimple Lie group and $\lg$ its Lie algebra. We denote by $\cal A$ the
space of (irreducible) connections on $P$,
and by $\cal G$ the infinite dimensional gauge group of vertical automorphisms
of $P$ (modulo the center of $G$). Then ${\cal G}$ acts freely on ${\cal A}$
and
\be
\Pi:{\cal A}\ra\C\label{65}
\ee
is a principal ${\cal G}$ bundle. The aim of this section will be to
determine a connection and curvature on this principal bundle, so that we
can write down (or recognize) the \mq\ form for some infinite dimensional
vector bundles associated to it. We will also state the Gauss-Codazzi
equation which express the Riemann curvature tensor $\RM$ of some
moduli subspace $\M$ of $\C$ in terms of the curvature of $\C$ and the
extrinsic curvature (second fundamental form) of the embedding
$\M\hookrightarrow\C$. The details can be found e.g.~in \cite{nr,ims,bv,gp}.

Continuing with notation, we denote by $\O{k}{M}$ the space of
$k$-forms on $M$ with values in the adjoint bundle $ad\,P:=P\times_{ad}\lg$
and by
\be
d_{A}: \O{k}{M}\ra\O{k+1}{M}\label{66}
\ee
the covariant exterior derivative with curvature $(d_{A})^{2}=F_{A}$.
The spaces $\O{k}{M}$
have natural scalar products defined by the metric $g$ on $M$ (and the
corresponding Hodge operator $*$) and an invariant scalar product $tr$ on
$\lg$, namely
\be
\langle X,Y \rangle = \int_{M}tr(X*Y)\;\;,\;\;\;\;\;\;\;\;X,Y\in\O{k}{M}
\label{67}
\ee
(I hope that occasionally denoting these forms by $X$ as well will not give
rise to any confusion with the manifold $X$ of section 2).
The tangent space $T_{A}{\cal A}$ to $\cal A$ at a connection $A$ can be
identified with $\O{1}{M}$ (as ${\cal A}$ is an affine space, two connections
differing by an element of $\O{1}{M}$).
Equation (\ref{67}) thus defines a metric $g_{\cal A}$ on $\cal A$.
The Lie algebra of ${\cal G}$ can be identified with $\O{0}{M}$ and acts on
$A\in{\cal A}$ via gauge transformations,
\be
A\mapsto A+d_{A}\Lambda\;\;,\;\;\;\;\;\;\;\;\;\;\Lambda\in\O{0}{M}
\label{68}\;\;,
\ee
so that $d_{A}\Lambda$ is the fundamental vector field at $A$ corresponding
to $\Lambda$. At each point $A\in\cal A$, $T_{A}{\cal A}$ can thus be
split into a vertical part $V_{A}={\rm Im}(d_{A})$ (tangent to the orbit of
$\cal G$ through $A$) and a horizontal part $H_{A}={\rm Ker}(d_{A}^{*})$ (the
orthogonal complement of $V_{A}$ with respect to the scalar product
(\ref{67})).
Explicitly this decomposition of $X\in\O{1}{M}$ into its vertical and
horizontal
parts is
\bea
X&=&d_{A}G_{A}^{0}d_{A}^{*}X + (X-d_{A}G_{A}^{0}d_{A}^{*}X)\;\;,\nonumber\\
 &\equiv& v_{A}X + h_{A}X\;\;,\label{69}
\eea
where $G_{A}^{0} = (d_{A}^{*}d_{A})^{-1}$ is the Greens function of the
scalar Laplacian (which exists if $A$ is irreducible). We will
identify the tangent space $T_{[A]}\C$ with $H_{A}$ for some representative
$A$ of the gauge equivalence class $[A]$.

Then $g_{\cal A}$ induces a metric $g_{\C}$ on $\C$ via
\be
g_{\C}([X],[Y])=g_{\cal A}(h_{A}X,h_{A}Y)\label{g1} \;\;,
\ee
where $X,Y\in\O{1}{N}$ project to $[X],[Y]\in T_{[A]}\C$.
With the same notation the Riemannian curvature of $\C$ is
\bea
\langle\RC([X],[Y])[Z],[W]\rangle&=&
\langle *[h_{A}X,*h_{A}W],G_{A}^{0}*[h_{A}Y,*h_{A}Z]\rangle -
(X\leftrightarrow Y)\nonumber\\
&+ & 2 \langle *[h_{A}{W},*h_{A}{Z}],G_{A}^{0}*[h_{A}{X},*h_{A}{Y}]\rangle
\label{g2}\;\;.
\eea
If $\M$ is some embedded submanifold of $\C$, then (\ref{g1}) induces a
metric $g_{\M}$ on $\M$ whose Riemann curvature tensor is
\bea
\langle\RM([X],[Y])[Z],[W]\rangle&=&\langle\RC([X],[Y])[Z],[W]\rangle
\nonumber\\
&+&(\langle K_{\M}([Y],[Z]),K_{\M}([X],[W])\rangle -(X\leftrightarrow Y))
\;\;,\label{g3}
\eea
where $K_{\M}$ is the extrinsic curvature (or second fundamental form)
of $\M$ in $\C$. For instanton moduli spaces $K_{\M}$ has been computed
in \cite{gp} and for moduli spaces of flat connections in two and
three dimensions one finds \cite{btmq}
\be
K_{\M}([X],[Y])=-d_{A}^{*}G_{A}^{2}[\bar{X}_{A},\bar{Y}_{A}]\label{g4}\;\;.
\ee
Here the tangent vectors $[X]$ and $[Y]$ to $\M$ are represented on
the right hand side by elements $\bar{X}$ and $\bar{Y}$ of $\O{1}{M}$
satisfying both the horizontality condition
$d_{A}^{*}\bar{X}=d_{A}^{*}\bar{Y}=0$ and the linearized flatness
equation $d_{A}\bar{X}=d_{A}\bar{Y}=0$. $G_{A}^{2}$ is the Greens
function of the Laplacian on two-forms and in the three-dimensional
case we think of it as being composed with a projector onto the
orthogonal complement of the zero modes of the Laplacian. Thus
\be
\langle K_{\M}([Y],[Z]),K_{\M}([X],[W])\rangle =
\langle [\bar{Y}_{A},\bar{Z}_{A}],G_{A}^{2}[\bar{X}_{A},\bar{W}_{A}]\rangle
\label{g5}
\ee
and together with (\ref{g2}) and (\ref{g3}) this determines $\RM$
entirely in terms of Greens functions of differential operators on $M$.
It is in this form that we will encounter $\RM$ in section 4.3.

The decomposition (\ref{69}) also
defines a connection on the principal bundle ${\cal A}\ra\C$ itself,
with connection form $\t_{{\cal A}}=G_{A}^{0}d_{A}^{*}$. Indeed,
$\t_{{\cal A}}$ can be regarded as a Lie algebra ($=\O{0}{M}$) valued
one-form on $\cal A$,
\bea
\t_{\cal A}: T_{A}{\cal A}&\ra&\O{0}{M}\nonumber\\
X&\mapsto& \t_{{\cal A}}(X)=G_{A}^{0}d_{A}^{*}X\label{70}\;\;.
\eea
It transforms homogenously under gauge transformations,
is obviously vertical (i.e.~vanishes on ${\rm Ker}(d_{A}^{*})$), and
assigns to the fundamental vector field $d_{A}\Lambda$ the corresponding
Lie algebra element
\be
\t_{\cal A}(d_{A}\Lambda) = G_{A}^{0}d_{A}^{*}d_{A}\Lambda = \Lambda
\label{71}\;\;,
\ee
as behoves a connection form. Its curvature is the horizontal two-form
\be
\Theta_{{\cal A}}=d_{\cal A}\t_{\cal A}+\frac{1}{2}[\t_{\cal A},\t_{\cal A}]
\label{72}
\ee
($d_{\cal A}$ denotes the exterior derivative on $\cal A$). Evaluated on
horizontal vectors $X,Y\in H_{A}$ the second term is zero
and from the first term only the variation of $A$ in $d_{A}^{*}$
will contribute (because otherwise the surviving $d_{A}^{*}$ will annihilate
either $X$ or $Y$). Thus one finds
\be
\Theta_{\cal A}(X,Y)=G_{A}^{0}*[X,*Y]\;\;,\label{73}
\ee
a formula that we will reencounter in our discussion of Donaldson theory
below.

Finally, we will introduce the bundles ${\cal E}_{0}$ and ${\cal E}_{+}$
which will play a role in the interpretation of topological gauge
theories from the \mq\ point of view below.
If $\dim(M)=2$, we consider the bundle
\be
{\cal E}_{0}:={\cal A}\times_{\cal G}\O{0}{M} \label{e0}
\ee
associated to the principal bundle (\ref{65}) via the adjoint representation.
If $\dim(M)=4$, we choose as fibre the space $\Omega^{2}_{+}(M,\lg)$
of self-dual two-forms. One then has the associated vector bundle
\be
{\cal E}_{+}:={\cal A}\times_{\cal G}\Omega^{2}_{+}(M,\lg) \label{e+}
\ee
over $\C$. In the standard manner (\ref{e0}) and (\ref{e+}) inherit
the connection (\ref{70}) and its curvature (\ref{73}) from the parent
principal bundle ${\cal A}\ra\C$ (\ref{65}).

\subsection{The Atiyah-Jeffrey Interpretation of Donaldson theory}

Donaldson theory \cite{ewdon} is the prime example of a cohomological
field theory. It was introduced by Witten to give a field theoretic
description of the intersection numbers of moduli spaces of instantons
investigated by Donaldson \cite{don}. Donaldson's introduction
of gauge theoretic methods into the study of four-manifolds has had enormous
impact on the subject (see \cite{fourman} for reviews), but unfortunately
it would require a seperate set of lectures to describe at least the basic
ideas. Likewise, it is not possible to give an account of the field theoretic
description here which would do justice to the many things that can and
should be said about Donaldson theory. Therefore, I will make only a few
general remarks on the structure of the action of Donaldson theory and
other cohomological field theories describing intersection theory on moduli
spaces. The main aim of this section will,
of course, be to show that this action is, despite appearance, also of the
\mq\ type. For a review of both the mathematical
and the physical side of the story see \cite[pp.~198-247]{pr}.

The action of Donaldson theory on a four-manifold $M$ in equivariant form
(i.e.~prior to the introduction of gauge ghosts) is \cite{ewdon}
\bea
S_{D}&=&
\int_{M}\left(B_{+}(F_{A})_{+} +\c_{+}(d_{A}\p)_{+} -\a B_{+}^{2}/2 +
\e d_{A}*\p\right)
\nonumber\\
 & &+ \left(\fb d_{A}*d_{A}\f + \fb[\p,*\p] -\a\f [\c_{+},\c_{+}]/2\right)
\;\;.\label{75}
\eea
Here $(.)_{+}$ denotes projection onto the self-dual part of a two-form,
\be
(F_{A})_{+}=\frac{1}{2}(F_{A}+*F_{A})\;\;,\;\;\;\;\;\;*(F_{A})_{+}=(F_{A})_{+}
\;\;,\label{76}
\ee
etc. Furthermore $\p\in\O{1}{M}$ is a Grassmann odd Lie algebra valued
one-form with ghost number 1. It is (as in \sqm) the superpartner of the
fundamental bosonic variable $A$ and represents tangent vectors to $\cal A$.
$(B_{+},\c_{+})$ are self-dual
two-forms with ghost numbers $(0,-1)$ (Grassmann parity (even,odd)),
and $(\f,\fb,\eta)$ are elements of $\O{0}{M}$ with ghost numbers $(2,-2,-1)$
and parity (even,even,odd). $\a$ is a real parameter whose significance is
the same as that played by $\a$ in \sqm\ (cf.~(\ref{58})).
This action has an equivariantly nilpotent BRST-like symmetry
\bea
\d A &=& \p\;\;\;\;\;\;\;\;\;\;\d \p = -d_{A}\f\nonumber\\
\d \c_{+}&=& B_{+}\;\;\;\;\;\d B_{+} = [\f,\c_{+}]\nonumber\\
\d \fb &=& \eta\;\;\;\;\;\;\;\;\;\;\;\d \eta = [\f,\fb]\nonumber\\
\d \f &=& 0 \;\;\;\;\;\;\;\;\;\;\;\d^{2}=\d_{\f}\label{77}
\eea
where $\d_{\f}$ denotes a gauge variation with respect to $\f$.
{}From these transformations it can be seen that the action $S_{D}$ is
BRST-exact,
\be
S_{D}=\d\int_{M}\c_{+}((F_{A})_{+}-\a B_{+}/2) + \fb d_{A}*\p\;\;.\label{78}
\ee
(cf.~(\ref{exact},\ref{54})).
The single most important consequence of (\ref{78}), which we will
abbreviate to $S_{D}=\d\Sigma_{D}$, is that the partition
function $Z(S_{D})$ of $S_{D}$ is given exactly by its one-loop
approximation. Likewise, it is independent of the metric on $M$ and any
other `coupling constants' which may enter into its construction
in addition to $\hbar$ and $g_{\m\n}$.
E.g.~for the metric the argument runs as
follows. Although $g_{\m\n}$
enters in a number of places in (\ref{75}), a variation
of it produces an insertion of a BRST-exact operator into the path
integral whose vacuum expectation value vanishes provided that the
vacuum is BRST invariant,
\bea
{\textstyle\frac{\d}{\d g_{\m\n}}}Z(S_{D})&=&
{\textstyle\frac{\d}{\d g_{\m\n}}}\int e^{-\d\Sigma_{D}}\nonumber\\
&=&-\langle 0|\d({\textstyle\frac{\d}{\d g_{\m\n}}}\Sigma_{D})|0\rangle=0
\label{79}\;\;.
\eea
By the same argument, $Z(S_{D})$ is independent of $\a$ and correlation
functions of metric independent and BRST invariant operators are
themselves metric independent. We will briefly come back to these
`observables' of Donaldson theory below.

Equation (\ref{78}) also makes the significance of the individual terms
in (\ref{75}) more transparent. In particular,
one sees that the first term of (\ref{78}) imposes
a delta function ($\a =0$) or Gaussian (for $\a\neq 0$)
constraint onto the instanton configurations $(F_{A})_{+}=0$. Together
with the gauge fixing of the gauge fields $A$, implicit in the above,
this localizes the path integral around the instanton moduli space
$\M_{I}$. The second
term, on the other hand, fixes the tangent vector $\p$ to be horizontal,
i.e.~to satisfy $d_{A}^{*}\p=0$, and $\p$ thus represents a tangent vector
to $\C$. Moreover, the $\c_{+}$ equation of motion restricts $\p$ further
to be tangent to $\M_{I}$, i.e.~to satisfy the linearized
instanton equation $(d_{A}\p)_{+}=0$ (modulo irrelevant terms proportional
to $\a$). The number of $\p$ zero modes will thus (generically, see
\cite{ewdon,pr}) be equal to the dimension $d(\M)$ of $\M_{I}$.

The structure of Donaldson theory summarized in the preceding paragraphs
is prototypical for the actions of cohomological field theories in general:
Given the moduli space $\M$ of interest, one seeks a description of it in
terms of certain fields (e.g.~connections), field equations
(e.g.~$(F_{A})_{+}=0$), and their symmetries (e.g.~gauge symmetries). One
then constructs an action which is essentially a bunch of delta functions
or Gaussians around the desired field configurations and (by supersymmetry)
their tangents. Thus, a topological action describing intersection theory
on the moduli space of flat connections on some $n$-manifold $M$ would roughly
be of the form
\be
S\sim\int_{M}
B_{n-2}F_{A}+ {\rm (super\;\;partners)} + {\rm (gauge\;\;fixing\;\;terms)}\;\;,
\label{80}
\ee
where $B\in\O{n-2}{M}$ and (for the cognoscenti) `gauge fixing terms'
is meant to also include all the terms corresponding to the higher
cohomology groups of the deformation complex of $\M$, i.e.~to the tower
of Bianchi symmetries $\d_{B}B_{n-2}=d_{A}B_{n-3},\d_{B}B_{n-3}=\ldots$.

Evidently, this is quite a pragmatic and not very
sophisticated way of looking at \tft.
It will, however, be good enough for the time being. Later on we will see
how to construct the action (\ref{80}) from the more satisfactory
\mq\ point of view. For an elaboration of the axiomatic approach
initiated by Atiyah \cite{at2} see \cite[chs.~3 and 4]{scott}.

Let us now return to Donaldson theory and show that its action $S_{D}$ is
of the \mq\ form. We will do this by making use of the equations of motion
arising from (\ref{78}) (which is legitimate since all the integrals are
Gaussian). We set $\a=1$ in the following.
\begin{itemize}
\item Integrating out $B$ one obtains the term $-(F_{A})_{+}^{2}/2$
\item The $\eta$-equation implies that $\p$ is horizontal which is
      henceforth tacitly understood
\item The $\fb$ equation of motion yields
\be
\f=G_{A}^{0}*[\p,*\p]\;\;,\label{81}
\ee
and, plugged back into the action, this gives rise to the term
\[-[\c_{+},\c_{+}]G_{A}^{0}[\p,*\p]/2\;\;.\]
\item Putting all this together we see that effectively the action of
Donaldson theory is
\be
S_{D}=\int_{M} -(F_{A})_{+}^{2}/2
-[\c_{+},\c_{+}]G_{A}^{0}[\p,*\p]/2 + (d_{A}\p)_{+}\c_{+}\;\;.\label{82}
\ee
\end{itemize}
Let us now compare this with (\ref{28}). We see that, apart from a factor of
$i$ which is not terribly important and which can be smuggled back into
(\ref{75}) and (\ref{82}) by appropriate scaling of the fields), the
correspondence is perfect. From the identification $\c_{a}\sim\c_{+}$ we
read off that the standard fibre of the sought for vector bundle is
$\Omega^{2}_{+}(M,\lg)$. The section is obviously $s(A)=(F_{A})_{+}$,
and as this transforms in the adjoint under gauge transformations the
vector bundle in question has to be the bundle ${\cal E}_{+}$ introduced
in (\ref{e+}). This is also confirmed by a comparison of the second term
of (\ref{28}) with the second term of (\ref{82}) and the curvature form
$\Theta_{\cal A}$ (\ref{73}). Thus we finally arrive at the desired
equation \cite{aj}
\be
Z(S_{D})=\cs({\cal E}_{+})\label{83}
\ee
identifying the partition function of Donaldson theory as the regularized
Euler number of the infinite dimensional vector bundle ${\cal E}_{+}$
and proving the result claimed in example 2 of section 2.3.

One important point we have ignored so far is that the partition function
$Z(S_{D})$ will be zero whenever there are $\p$ zero modes, i.e.~whenever
the dimension $d(\M)$ of $\M_{I}$ is non-zero. This is in marked
contrast with the situation we encountered in \sqm\ in section 3.
There the partition function $Z(S_{M})=\c(M)$ was generally non-zero,
despite the presence of $\dim(M)$ $\p$ zero modes. I will now briefly
try to explain the reason for this difference and the related issue
of observables in Donaldson theory (with no claim to completeness nor
to complete comprehensibility):

In \sqm\ there are an equal number of $\p$ and $\pb$ zero modes, and
these can be soaked up by expanding the curvature term (which contains
an equal number of $\p$'s and $\pb$'s) to the appropriate power. In
Donaldson theory the role of $\pb$ is played by $\c_{+}$. Generically,
however, there will be no $\c_{+}$ zero modes at all, independently
of the dimension of the moduli space, so that the fermionic $\p$
zero modes can not be soaked up by the curvature term of
(\ref{82}). (As an aside: the $\c_{+}$ zero modes represent the
second cohomology group of the instanton deformation complex and thus,
together with reducible connections, the obstruction to having a
smooth moduli space. For the class of four-manifolds considered in
\cite{don} it can be shown that this cohomology group is zero at
irreducible instantons for a generic metric.)

Thus, in order to get a non-zero result one has to insert operators
into the path integral which take care of the $\p$ zero modes or,
in other words, one has to construct a top-form on $\M_{I}$
which can then be integrated over it. These operators have to be
BRST invariant, and - in view of (\ref{77}) - this translates into the
requirement that they represent cohomology classes of $\C$.
This is just like the situation we considered at the end of section 2.
When there is a mismatch between the rank $2m$ of $E$ and the dimension
$n$ of $X$ one can obtain non-zero numbers by pairing $\en(E)$ with
representatives of $H^{n-2m}(X)$. Likewise, even if $n=2m$ but one
chooses a non-generic
section of $E$ with a $k$-dimensional zero locus, this can be
represented by an $(n-k)$-form which still has to be paired with a
$k$-form in order to make it a volume form on $X$. In the case of
Donaldson theory we have chosen a section with a $d(\M)$-dimensional
zero locus and we have to pair the corresponding Euler class,
the integrand of (\ref{83}), with $d(\M)$-forms on $\C$ to produce a
good volume form on $\C$ which will then localize to a volume form on
$\M_{I}$. In the work of Donaldson the cohomology classes considered
for this purpose are certain characteristic classes (of the universal
bundle of \cite{as}) which also arise naturally in the field theoretic
description \cite{bs,pr}. For instance, one of the building blocks is
the two-form $\f$ as given by
(\ref{81}) which  represents the curvature form $\Theta_{\cal A}$
(\ref{73}). Unfortunately, these intersection numbers are very difficult
to calculate in general. For details please consult the cited literature.

\subsection{Flat connections in two and three dimensions}

It is, of course, also possible to turn around the strategy of the
previous section, i.e.~to start with the \mq\ formalism applied
to some vector bundle over $\C$ and to then reconstruct the action
of the corresponding \tgt\ from there.

Let us, for instance, consider
the problem of constructing a \tgt\ in $3d$ whose partition function
(formally) calculates the Euler characteristic $\c(\M^{3})$ of the
moduli space $\M^{3}=\M^{3}(M,G)$ of flat $G$ connections on some
three-manifold $M$. We actually already know two ways of achieving
this, provided that we can find a vector field $v$ on $\A{3}$ (the
superscripts are a reminder of the dimension we are in) whose
zero locus is $\M^{3}$. Fortuitously, in three dimensions such
a vector field exists, namely $v=*F_{A}$. A priori, this only defines
a vector field on ${\cal A}^{3}$, as $*F_{A}\in\O{1}{M}$. It is, however,
horizontal ($d_{A}^{*}*F_{A}=0$ by the Bianchi identity $d_{A}F_{A}=0$)
and thus projects to a vector field on $\A{3}$ whose zero locus is
$\M^{3}$. This vector field is the gradient vector field of the
Chern-Simons functional
\be
CS(A)=\int_{M}AdA +{\textstyle\frac{2}{3}}A^{3}\label{84}\;\;,
\ee
whose critical points are well known to be the flat connections.
(Of course, this does not really define a functional on $\A{3}$,
as it changes by a constant times the winding number under large
gauge transformations. But its derivative is well defined and
this non-invariance implies that the one-form $d_{\cal A}CS(A)$
passes down to a closed but not exact one-form ${\cal F}_{A}$
on $\A{3}$. Explicitly, ${\cal F}_{A}$ is given by
\bea
{\cal F}_{A}:T_{[A]}\A{3}&\ra&{\bf R}\nonumber\\
                {[X]}&\ra&\int_{M}F_{A}X\label{85}\;\;.
\eea
Note that this does not depend on the representative of $[X]$ as
$\int_{M}F_{A}d_{A}\Lambda=0$.) In two dimensions such a vector
field appears not to exist at first sight
and one has to be a little more inventive (cf.~\cite{btmq} and the
remarks at the end of this section).

Given this vector field, the first possibility is then to adapt the
Atiyah-Jeffrey construction of the previous section to the case
$X=\A{3}$ and $E=T(\A{3})$, to use $v=*F_{A}$ as the regularizing
section for
\be
\c_{v}(\A{3})= \c(\M^{3})\label{86}\;\;,
\ee
and to represent this by the functional integral
\be
\c(\M^{3})=\int_{\A{3}}e_{v,\nabla}(\A{3})\label{87}\;\;.
\ee
Of course, the `action', i.e.~the exponent of (\ref{87}), will contain
non-local terms like the curvature tensor $\RC$ (\ref{g2}), as in
(\ref{82}). As this is undesirable for a fundamental action, we will
introduce auxiliary fields (like those we eliminated in going
from (\ref{75}) to (\ref{82})) to rewrite the action in local form.

Alternatively, we can construct \sqm\ on $\A{3}$ using $\dot{A}+v$
as the section of $T(L\A{3})$, i.e.~we use the action $S_{M,W}$
(\ref{58}) of section 3 and substitute $M\ra\A{3}$ and $W\ra CS(A)$.
This will give us a (non-covariant) $(3+1)$-dimensional gauge theory
on $M\times\s1$ (in fact, the $(3+1)$-decomposition of Donaldson theory,
see \cite{at,ewdon} and \cite{btqm} for details).
However, from the general arguments of section 3 we know
that only the constant Fourier modes will contribute, so that one is
left with an effective three-dimensional action which is identical
to the one obtained by the first method.

Irrespective of how one
chooses to go about constructing the action (there are still further
possibilities, see e.g.~\cite{ewtop,bbt,btmq}), it reads
\bea
S_{\M}&=&\int_{M}\left(B_{1}F_{A}+\a B_{1}*B_{1}/2 + d_{A}u*d_{A}u/2
-d_{A}\fb*d_{A}\f +\pb d_{A}\p\right)\nonumber\\
&+&\left(u[\p,*\pb]+ \e d_{A}*\p + \eb d_{A}*\pb +\fb[\p,*\p] -
\a\f[\pb,*\pb]/2\right)\label{88}\;\;.
\eea
$u$ is a scalar field, and as in \sqm\ we have denoted the field $\c$ of the
\mq\ formula by $\pb$. The rest should look familiar. Superficially,
this action is very similar to that of Donaldson theory. There is a
Gaussian constraint onto flat connections, the tangents $\p$ have to
satisfy the linearized flatness equations, and there are cubic
interaction terms involving the scalar fields $\f,\fb$ and $u$.
However, there is one important difference, namely that there is a perfect
symmetry between $\p$ and $\pb$. As in \sqm, both represent tangent
vectors, we also see that both are gauge fixed to be horizontal,
and both have to be tangent to $\M^{3}$. In particular, therefore,
there will be an equal number of $\p$ and $\pb$ zero modes and we
have the possibility of obtaining a non-zero result even if
$\dim(\M^{3})\neq 0$. This is reassuring as we, after all, expect to
find $Z(S_{\M})=\c(\M^{3})$. Let us now show that this is indeed the case.
\begin{itemize}
\item First of all integration over $\e$ and $\eb$ forces $\p$ and
      $\pb$ to be horizontal, $h_{A}\p=\p$, $h_{A}\pb=\pb$,
      i.e.~to represent tangent vectors to $\A{3}$
\item Setting $\a =1$, integration over $\fb$ yields
      $\f=-G_{A}^{0}*[\p,*\p]$, giving rise to a term
      \[\langle *[\pb,*\pb],G_{A}^{0}*[\p,*\p]\rangle/2\]
      in the action
\item The equation of motion for $u$ reads
      \[u=G_{A}^{0}*[\p,*\pb]\]
      and plugging this back into the action one obtains a term
 \[\langle *[\p,*\pb],G_{A}^{0}*[\p,*\pb]\rangle/2\]
\item This combination of Greens function is precisely that appearing
      in the formula (\ref{g2}) for the Riemann curvature tensor
      $\RC$.Thus we have already reduced the action to the form
      $S_{\M}=\RC + {\rm `something'}$ and we expect the `something'
      to be the contribution (\ref{g5}) to $\RM$ (\ref{g3}) quadratic
      in the extrinsic curvature $K_{\M}$.
\item To evaluate the integral over the remaining fields $A$, $\p$, and
      $\pb$ we expand them about their classical configurations which
      we can take to be flat connections $A_{c}$ and their tangents
      (because of $\a$-independence). By standard arguments we may
      restrict ourselves to a one-loop approximation and to this order
      the remaining terms in the action become
\[\int_{M}(d_{A_{c}}A_{q}*d_{A_{c}}A_{q}/2 + [\pb_{c},\p_{c}]A_{q})\;\;.\]
\item Finally, integration over $A_{q}$ yields
\[\langle [\pb_{c},\p_{c}],G_{A_{c}}^{2}[\pb_{c},\p_{c}]\rangle/2\;\;,\]
      which we recognize to be precisely the contribution
      (\ref{g5}) of $K_{\M}$. Thus we have reduced the action (\ref{88})
      to $\RM$, expressed in terms of the classical configurations
      $A_{c}$, $\p_{c}$ and $\pb_{c}$. We are now on familiar ground
      (see e.g.~(\ref{25},\ref{fred})) and know that evaluation of this
      finite dimensional integral gives
\be
      Z(S_{\M})=\c(\M)\label{89}\;\;.
\ee
\end{itemize}
This calculation also illustrates how the Gauss-Codazzi equations emerge
from the Mathai-Quillen form in general. Guided by this example it is
now straightforward to perform the analogous manipulations in the finite
dimensional case (section 2) and in \sqm\ (section 3).

We end this $3d$ example with the remark that,
by a result of Taubes \cite{tau}, the partition function
of (\ref{24}) formally equals the {\em Casson invariant} of $M$ if $M$ is
a homology three-sphere \cite{ewtop}. This, combined with the above
considerations, has led us to propose $\c(\M)$ as a candidate for the
definition of the Casson invariant of more general three-manifolds
(see \cite{btmq} for some preliminary considerations).

The simplest example to consider in two dimensions is the
analogue of Donaldson theory, i.e. a \tft\ describing
intersection theory on a moduli space $\M^{2}$ of flat connections
in two dimensions. Instead of the bundle ${\cal E}_{+}$ with standard
fibre $\Omega^{2}_{+}(M,\lg)$ (\ref{e+}) we choose the bundle
${\cal E}_{0}$ (\ref{e0}) with standard fibre $\O{0}{M}$.
This will have the effect of replacing
the self-dual two-forms $B_{+}$ and $\c_{+}$ of Donaldson theory
by zero-forms $B_{0}$ and $\c_{0}$. A natural section of ${\cal E}_{0}$
is $s(A)=*F_{A}$ with zero locus $\M^{2}$. This results in the
trading of $(F_{A})_{+}$ and its linearization $(d_{A}\p)_{+}$ for
$F_{A}$ and its linearization $d_{A}\p$ in the action (\ref{75}).
With this dictionary in mind the action is precisely the same as
that of Donaldson theory. It is also the $2d$ version of (\ref{80})
and we have thus just completed the construction of\\
\noindent{\bf Example 5}
\underline{$X=\A{2}$, $E={\cal E}_{0}$, $s=*F_{A}$}\\
The fundamental reason for why this theory is so similar to
Donaldson theory is that in both cases the deformation complex
is short so that one will find essentially the same field content.
In three dimensions, on the other hand, the deformation complex
is longer by one term and this is reflected in the appearance of the scalar
field $u$ in (\ref{88}).

Again, the partition function, i.e.~the
regularized Euler number of ${\cal E}_{0}$, will vanish when
$\dim(\M^{2})\neq 0$.
But, none too surprisingly, there also exist analogues of the Donaldson
polynomials, the observables of Donaldson theory, which come to the
rescue in this case. Life in two dimensions is easier than in four,
and the corresponding intersection
numbers have indeed been calculated recently by Thaddeus \cite{thadd}
using powerful tools of conformal field theory and algebraic geometry
(see also \cite{ewym,donbon}).

As our final example let us consider a topological gauge theory
representing the Euler characteristic of $\M^{2}$. As mentioned above,
$*F_{A}$ is not a vector field on $\A{2}$, so that it is not immediately
obvious which section of $T\A{2}$ to choose. The dimensional reduction
of the action (\ref{88}) suggests, that the right base space to consider
is $X={\cal A}^{2}\times\O{0}{M}$, where the second factor represents the
third component $\rho$ of $A$. Then a possible section of $TX$ is
$V(A,\rho)=(*d_{A}\rho, *F_{A})$ whose zero locus (for irreducible $A$)
is indeed precisely the space of flat connections. But this is not the
complete story yet. The problem is, that $*d_{A}\rho$ is only horizontal
if $A$ is flat ($d_{A}^{*}*d_{A}\rho = [*F_{A},\rho]$). Thus,
one possibility is to use a delta function instead of a Gaussian
constraint onto flat connections ($\a=0$). This action can be found in
\cite{btmq}. Alternatively, one might attempt to replace $*d_{A}\rho$ by
$h_{A}*d_{A}\rho$. This necessitates the introduction of additional
auxiliary fields to eliminate the non-locality of $h_{A}$, and a more
detailed investigation of this possibility is left to the reader.

\subsubsection*{Acknowledgements}

I wish to thank R.~Gielerak and the whole organizing committee for
inviting me to lecture at this School. Thanks are also due to all the
participants for creating such a stimulating atmosphere,
and to Jaap  Kalkman for sending me \cite{jk}. Finally, I wish to
acknowledge the financial support of the Stichting FOM.

\rnc{\Large}{\normalsize}

\end{document}